\documentclass{article}
\usepackage[a4paper, total={6in, 9in}]{geometry}
\usepackage{amsmath}
\usepackage{xcolor}
\usepackage{graphicx}
\usepackage{multirow}
\usepackage{url}
\usepackage[authoryear,round]{natbib}
\usepackage{authblk}

\title{Normative boundaries of AI in scientific work: Evidence from PhD researchers}

\author[1]{Francesco Angelini\thanks{Corresponding author. email: francesco.angelini7@unibo.it}}
\author[2]{Johan Lyrvall}

\affil[1]{Independent Researcher, Italy}
\affil[2]{National Institute for Research in Digital Science and Technology (Inria), Villeneuve-d'Ascq, France}

\begin{document}
\maketitle

\begin{abstract}

Artificial intelligence (AI) is increasingly embedded in scientific work, but researchers may not evaluate its use uniformly across research tasks. This study examines task-specific attitudes towards AI among an international, self-selected sample of 3,785 PhD students in STEM and medical and health sciences who participated in Nature’s Graduate Survey 2025. We analyse respondents' comfort with using AI for writing a research article, collecting and analysing data, designing experiments, tracking scientific literature, and summarising it. Latent class analysis identifies four distinct attitudinal profiles. The dominant profile reflects a ``division of labour,'' in which AI is widely accepted for literature-related tasks but resisted in activities closely associated with intellectual contribution, such as writing, data analysis, and experimental design. A ``status quo'' profile is broadly uncomfortable across tasks, an ``all-purpose'' profile is broadly comfortable, and an ``undecided'' profile expresses substantial uncertainty. These patterns suggest that attitudes towards AI in research are organised less around a simple acceptance--rejection divide than around task-specific boundaries, likely concerning delegation, authorship, and responsibility. Because the survey measures comfort rather than legitimacy, the profiles are best interpreted as attitudinal configurations with a normative dimension. The findings highlight the importance of task-specific approaches to AI governance, doctoral training, disclosure, and research evaluation.
\end{abstract}


\section{Introduction}\label{sec:intro}
The rapid diffusion of artificial intelligence (AI), particularly generative AI tools, is reshaping not only how work is performed, but also how tasks are allocated between humans and machines \citep{eloundou2024}. Evidence from knowledge-intensive occupations indicates that this interaction is highly uneven. AI has improved speed and output quality in professional writing tasks \citep{noy2023}, produced particularly large productivity gains among less experienced customer-support workers \citep{brynjolfsson2025}, and enhanced performance on some management-consulting tasks while reducing performance on others \citep{dellacqua2026}. These findings suggest that AI does not uniformly automate entire occupations or set of skills. Rather, it transforms the composition of work, complementing human expertise in some activities while substituting for, or altering the role of, human effort and judgement in others.

Scientific work is no exception to this task-level transformation. Recent work has conceptualised AI as a general method of invention, capable of reshaping how knowledge is generated and recombined \citep{bianchini2022}. Emerging empirical evidence also suggests that AI adoption is already affecting research productivity, knowledge discovery, and the signals used to evaluate scientific quality \citep{kusumegi2025}. However, these effects are unlikely to depend only on the availability or technical capabilities of AI. They also depend on which parts of the research process are delegated to AI, how human expertise is reorganised around those activities, and the norms that govern its use. 

These emerging normative boundaries are the central concern of this study. Survey evidence indicates that researchers see considerable potential in AI for scientific work, while also expressing concerns about bias, misuse, and reliability \citep{vannoorden2023}. We extend this perspective to the task level, examining how researchers evaluate AI use across different research activities and where they draw boundaries around acceptable forms of AI assistance. Such evaluations are not only abstract ethical positions. They may influence which research activities are considered legitimately delegable to AI, which remain associated with human authorship, and how scientific work is consequently organised. These questions are particularly salient for PhD students, who are encountering AI technologies at a formative stage of their careers, while norms about legitimate research practices are still being learned and negotiated. They are today users of AI tools and, at the same time, future researchers whose emerging attitudes may shape the practices through which AI becomes embedded in scientific work.

In practice, scientific work consists of multiple tasks that differ in their epistemic role, their relation to authorship, and their contribution to knowledge production. Yet, most existing discussions of AI in science focus either on aggregate adoption or on general attitudes toward its use. Such perspectives risk obscuring systematic heterogeneity in how researchers evaluate AI across different research activities. In this paper, we argue that understanding the role of AI in science requires moving beyond aggregate measures of adoption and examining how its use is differentiated across research tasks. Using survey data from Nature’s Graduate Survey 2025, we identify distinct attitudinal profiles that capture task-specific patterns of comfort with AI use and reflect broader normative differentiation across research activities.

Our findings suggest that AI is unlikely to become embedded in scientific work through uniform adoption. Rather, its integration appears to be structured by task-specific normative boundaries separating uses perceived as acceptable from those perceived as unacceptable. By documenting these emerging normative patterns among future researchers, this paper contributes to understanding how digital technologies become institutionalised within scientific practice and how they may reshape the division of labour in research. We also discuss the implications of these patterns for the present governance of AI in science, particularly for doctoral training, disclosure requirements, and research evaluation. Looking ahead, we consider how these boundaries may evolve as AI becomes increasingly normalised within research workflows, and how the attitudes documented among current PhD students may influence future research practices and the organisation of scientific work.

\section{Related literature and positioning}


Evidence across different domains suggests that evaluations of AI depend on the characteristics of the activity in which it is used and on the role assigned to the technology relative to human judgement. \citet{castelo2019} show that people are less willing to rely on algorithms for tasks perceived as subjective than for tasks perceived as objective. In healthcare, \citet{longoni2019} find that resistance to medical AI decreases when the technology is presented as supporting, rather than replacing, decisions made by human professionals. Although these studies do not concern scientific research, they establish a relevant premise: acceptance of AI assistance depends on its perceived technical capabilities, on the nature of the task, and on how responsibility is allocated between humans and machines.

Within academic contexts, research has primarily examined the factors associated with general AI assistance and use. A systematic review by \citet{acostaenriquez2024} organises this literature through the UTAUT2 framework, highlighting the role of expected performance, ease of use, social influence, facilitating conditions, and individual experience, while also identifying ethics and privacy as concerns specific to AI adoption in university settings. Empirical evidence similarly associates AI appropriation among university students with performance expectancy, social influence, and confidence in learning about AI \citep{acostaenriquez2025}. Among university researchers, \citet{acostaenriquez2025academia} find that social influence, educational self-efficacy, and academic integrity are associated with researchers' use of AI. Taken together, this literature shows that AI adoption in academia is shaped by technical, individual, social, and normative factors. It generally treats acceptance or use as an overall outcome, however, rather than examining how evaluations vary across specific research activities.

A more task-specific perspective is offered by research examining how AI is used across different components of scientific work. \citet{chugunova2026} survey researchers working in two major German research organisations and show that AI tools are already used across a broad range of research activities, including primary and creative tasks. Their analysis also points to the importance of familiarity and training, while identifying legal uncertainty and privacy concerns as relevant barriers to use. This study is particularly close to ours because it distinguishes between different research activities. Its primary focus, however, is on adoption, barriers, and perceived impacts, rather than on how researchers evaluate the acceptability of AI use across tasks.

A parallel literature examines the consequences of AI adoption on scientific outcomes. Recent evidence shows that its impact on research novelty and productivity is heterogeneous across contexts \citep{bianchini2026}. Other work documents changes in submission volumes and writing quality associated with the diffusion of generative AI, while deliberately avoiding a normative judgement about which uses should be considered appropriate \citep{gartenberg2026}. These studies provide important evidence on how AI may transform scientific production, but leave open the question of how researchers themselves evaluate different forms of AI assistance.

Qualitative evidence from doctoral work suggests that such evaluations may vary according to the role assigned to AI. Drawing from interviews with L2 doctoral students, \citet{hoomanfard2025} document the use of generative AI for language support, conceptual clarification, summarisation, and other dissertation-writing activities, alongside uncertainty about the boundary between legitimate assistance and plagiarism and about whether AI use should be disclosed. However, their analysis is centred specifically on second-language dissertation writing, rather than on the broader range of activities involved in scientific research. We extend this perspective by examining task-specific evaluations of AI among PhD researchers in STEM and medical and health sciences.


Studying these evaluations is relevant because attitudes form part of the context in which technology adoption occurs. Research on technology acceptance indicates that perceived benefits, social expectations, and facilitating conditions can shape individuals' willingness to use a technology \citep{venkatesh2021}. Attitudes should not, however, be treated as direct measures of behaviour. Researchers may use tools they evaluate negatively, or refrain from uses they regard as acceptable, because of differences in access, competence, professional incentives, or institutional constraints. Nevertheless, task-specific evaluations can suggest which forms of AI assistance researchers are more willing to accept, resist, or approach with uncertainty.

Taken together, existing research documents both widespread AI use and ambivalence regarding its appropriate application. Some studies already distinguish between research activities, but do not provide a systematic account of how norms governing AI use vary across research tasks. This distinction matters because aggregate measures may hide substantially different configurations of acceptance, resistance, and uncertainty. Mapping these configurations is also relevant for research governance, since institutional responses that treat AI use as a homogeneous practice may overlook important differences between literature-related assistance, writing, data analysis, and experimental design.

Against this background, we examine whether evaluations of AI use across core research activities form distinct and coherent attitudinal profiles among PhD researchers in STEM and medical and health sciences. This individual-level perspective allows us to move beyond both aggregate measures of acceptance and separate comparisons between tasks, and to investigate how different evaluations of AI use are combined within researchers.

\section{Method}
We use data from Nature’s Graduate Survey 2025 \citep{ngs2025}, conducted by Springer Nature in collaboration with Thinks Insights \& Strategy in May--June 2025. The survey collected responses from 3785 self-selected PhD students in hard-science fields across 107 countries, recruited via the \emph{Nature} journal website, other Springer Nature digital products, and targeted email campaigns.

Our analysis focuses on five survey items measuring respondents’ comfort with using AI tools for specific research activities: (i) writing a research article, (ii) collecting and analysing data, (iii) designing experiments, (iv) tracking scientific literature, and (v) summarising scientific literature. Each item has six response categories ranging from \emph{very comfortable} to \emph{very uncomfortable}, plus a \emph{don’t know} option. Specifically, the prompt for these items is \emph{How comfortable, if at all, are you with using AI tools (e.g. ChatGPT, Copilot and Claude), for each of the following activities that you may do as part of your PhD/research?} These items are particularly suited to capturing task-specific normative attitudes rather than general AI adoption.
The frequency distributions of the item responses are reported in Table~\ref{tab:summary_statistics_items}.
As can be seen, the PhD students lean negative on using AI for writing a research article, collecting and analysing data, and designing experiments, and lean positive on using AI for tracking and summarising scientific literature.
A notable proportion of PhD students have no opinion or have not yet made up their mind on using AI in their research.
These survey items have no missing values in the observed sample.

To identify the structure of normative attitudes across different research activities, we apply latent class analysis \citep[e.g.,][]{goodman1974} to these five variables.
Latent class analysis is a popular approach for identifying distinct profiles (classes) of individuals.
This works by assuming that the observed data has arisen from a mixture of underlying profiles, and ``back-tracking'' to uncover their unique characters.
In line with best practices \citep[e.g.][]{lyrvalletal2026}, we first select the number of latent classes by estimating and comparing a set of models.
Specifically, we start with a small number of latent classes and incrementally estimate more complex models until goodness-of-fit stops improving.
As decision criterion we consider the ICLbic \citep{biernackiceleuxgovaert2000}; an adjustment of the classical BIC for class separation.
In essence, the ICLbic is used to find an optimal trade-off between fit, complexity, and profile distinctiveness.
Subsequent to selecting an adequate model, we explore whether we can predict AI use attitudes based on academic characteristics.

We consider seven variables as potential predictors of AI use attitudes.
Their frequency distributions are reported in Table~\ref{tab:summary_statistics_covariates}.\footnote{The reported frequencies are based on the sample of 3722 respondents that have no missing values for any of the covariates, because these are the respondents that are used in the estimation that includes the covariates in Section~\ref{sec:results}.}
Most of these covariates are constructed from the original survey items by merging response categories.
The covariate indicating whether the respondent is a Native English speaker is constructed on the basis of their country of origin and the official United Kingdom list of nationalities that do not need to prove their knowledge of English.\footnote{
This list can be accessed at \url{https://www.gov.uk/english-language/exemptions}.
}
We can observe that the sample consists predominantly of students in science, technology, engineering, and mathematics (76\%).
The remaining respondents are studying for a PhD in medical and health sciences like medicine, public health, and pharmacy.
Early-stage PhD students in their first two years are a minority (38\%).
Most respondents study full-time (87\%).
AI use is widespread among the respondents: most (53\%) report using AI tools as part of their PhD research daily or weekly, and about a third (31\%) monthly or yearly.
PhD students of 34 years or younger constitute most of the sample (76\%).
A majority of the respondents reports being male (54\%), and a minority reports coming from a country that can be classified as native English speaking (15\%).

As is standard in latent class analysis \citep[e.g.][]{bakkkuha2018}, we integrate multinomial logistic regression into the latent class model.
The regression parameters can then be estimated conditionally on the profile characteristics uncovered in the previous step of the analysis.
We carry out all these steps of the quantitative analysis in the \texttt{R} package \texttt{multilevLCA} \citep{lyrvalletal2025}.

\section{Results}\label{sec:results}




As mentioned in the previous section, we select the number of latent classes based on the ICLbic information criterion, by starting low and incrementally increasing while goodness-of-fit improves.
Lower ICLbic values indicate better goodness-of-fit.
Table~\ref{tab:enumeration} reports the results.
As can be seen, the four-class model has the lowest ICLbic and locally optimal goodness-of-fit.
This can be interpreted as follows: the four-class model has the most preferable balance of fit, complexity, and profile distinctiveness.


The selected four-class model, reported in Table \ref{tab:fit_measurement_model}, reveals pronounced individual-level heterogeneity in attitudes towards AI use across research tasks.

The largest type (44\%), which we can label \emph{``division of labour''}, exhibits a selective pattern. These PhD students are relatively comfortable using AI for tracking and summarising scientific literature, but substantially less comfortable using AI for writing, data collection and analysis, and experiment design. The \emph{``division of labour''} type is consistent with a task-specific boundary between literature-oriented tasks and activities perceived as central to intellectual contribution and scientific responsibility.

In contrast, the smaller types exhibit consistent attitudes towards AI use across research activities.
The second-largest type (34\%), or \emph{``status quo''} in our taxonomy, is uncomfortable using AI in research.
However, the intensity of their attitudes varies across research activities, with stronger discomfort when it comes to writing, data collection and analysis, and experiment design, and milder discomfort when it comes to tracking and summarising literature.

The second-smallest type (16\%), or \emph{``all-purpose''}, broadly has great comfort using AI, suggesting that AI is viewed as a general-purpose research tool.
Despite their overall positive stance, the \emph{``all-purpose''} PhD student tends to have more positive attitudes towards using AI for tracking and summarising literature than for writing a research article, collecting and analysing data, and designing experiments.

The smallest type (7\%), which we label \emph{``undecided''}, is characterised by a high probability of selecting \emph{don’t know} responses for all research activities. This systematic response pattern should not necessarily be interpreted as a distinct attitudinal orientation. It may reflect uncertainty, limited familiarity or experience with AI tools and their application to specific tasks, non-use, or other response processes.
However, when the respondents in this class do express an attitude, it is more likely to be positive than negative for each of the research activities.

\subsection{Class prediction}

We fit the latent class model with covariates with the largest class (\emph{``division of labour''}) as reference category.
The estimates for the multinomial logistic parameters are reported in Table~\ref{tab:fit_structural_model}.
These were obtained by means of the two-step estimation approach \citep{bakkkuha2018} to guarantee that the identified class definitions reported in Table~\ref{tab:fit_measurement_model} are not distorted by the addition of the covariates to the model.

The fitted model suggests that there is variation in typical AI use attitudes across research fields.
We can observe a substantively and statistically significant estimate that PhD students in STEM are more likely than PhD students in medical and health sciences to be the \emph{``status quo''} type relative to the most frequent \emph{``division of labour''} type.
There is no statistical evidence on the 5\%-level that either of the remaining two types have a different composition of research fields relative to \textit{``division of labour.''}

There is also no evidence on the 5\%-level that typical AI use attitudes have different compositions of PhD students with respect to career stage (first two years of the PhD or later) or time commitment (full-time studies or not).
As such, considering that significantly research-aiding AI tools have been available for about two years, attitudes do not appear to have been shaped by greater availability of AI tools from the outset of the PhD.

PhD students using AI more frequently are, not surprisingly, estimated as being less likely to be the most skeptical \emph{``status quo''} type compared to \textit{``division of labour.''}
They are estimated as being still less likely to be \emph{``undecided''} in their attitudes towards using AI in their research.

\section{Discussion and conclusions}

Our findings suggest that attitudes towards AI use in doctoral research are not organised around a simple opposition between acceptance and rejection. Rather, they form task-specific configurations that are consistent with emerging normative boundaries around the delegation of research activities to AI. The dominant \emph{``division of labour''} profile is particularly informative in this respect: many PhD students appear willing to accept AI as a support tool for literature-related activities, while expressing less comfort with its use in activities more closely associated with authorship, creativity, and epistemic responsibility, such as writing, data collection and analysis, and experiment design.

These findings are also consistent with prior evidence that acceptance of algorithmic assistance depends both on the nature of the task and on the role assigned to the technology. \citet{castelo2019} show that reliance on algorithms varies across tasks and tends to be lower when activities are perceived as more subjective, while \citet{longoni2019} find that resistance to AI decreases when technology is presented as supporting rather than replacing the professional judgement. Our largest profile (\emph{``division of labour''}) points to that direction: literature tracking and summarisation can be seen as support tasks, in contrast with writing, designing experiments, and analysing data. The comparison must be interpreted cautiously: the survey data we use does not measure perceived task subjectivity. Our results therefore expand this literature by documenting task-specific configurations of AI acceptance within the scientific research process, rather than directly testing the mechanisms identified by the above-mentioned studies.

This result speaks to broader debates on the integration of AI into scientific production. If AI is increasingly becoming part of the research process, its integration is unlikely to occur uniformly across all stages of scientific work. Instead, our evidence suggests that researchers distinguish between tasks they are relatively comfortable delegating to AI and tasks that they may perceive as more closely tied to human scientific contribution. In this sense, AI adoption in science should not be understood only as a question of access, efficiency, or technical capability, but also as a matter of how researchers classify different forms of AI assistance as more or less legitimate.

\subsection{Implications for research governance and evaluation}
Our findings raise the question of whether concerns about AI use in research should focus on the process through which outputs are produced or on the quality of the resulting outputs. The task-specific boundaries we observe may suggest that researchers attach normative importance to how scientific work is carried out, not only to what is ultimately produced, but also to which activities are delegated to AI and which remain under direct human responsibility. A governance approach concerned exclusively with output quality may therefore overlook differences in authorship, accountability, and the distribution of intellectual contribution across the research process. At the same time, rules that treat any use of AI as a single and homogeneous practice may bring together substantially different forms of assistance.

Existing editorial policies already reflect, to some extent, this task-specific approach. Major publishers, to date, distinguish between AI use in conducting research, preparing a manuscript, and evaluating the work of others. These policies suggest that the relevant question is not whether AI was used, but for which activity, in what capacity, and under what degree of human control and supervision.

Our results suggest that this differentiation is also present in researchers' own evaluation. The dominant profile (\emph{``division of labour''}) distinguishes among assistance in literature-related tasks and assistance in experimental design, writing, and data analysis tasks, while the other profiles differ in whether they reject or accept AI assistance more broadly. Generic disclosure may then provide limited information if they do not specify the activity in which AI was used. More refined distinctions would make it easier to assess the allocation of responsibility without assuming that all AI-assisted activities carry the same epistemic or ethical implications.

Clearly, similar considerations apply to the evaluation of doctoral and scientific works. An outcome-based approach may treat an output as acceptable whenever it meets conventional standards of accuracy, novelty, quality, or other measures, regardless of how it was produced. A process-based approach, instead, may penalise AI assistance even when it is transparent, carefully supervised, and does not displace the researcher's substantive judgement. The task-specific patterns observed in our analysis suggest the need for a more differentiated approach, combining attention to output quality with transparency about the production process.

\subsection{AI, doctoral learning, and skill formation}
The task-specific boundaries we identified also raise questions about doctoral learning and skill development. Research activities performed during the PhD path have a dual function. They contribute to the production of the scientific outputs, but they are also activities through which doctoral researchers acquire substantive knowledge, methodological competence, and scientific judgement. The tasks considered in our analysis are therefore both activities whose completion AI may accelerate or partially automate and part of the process through which researchers learn to formulate arguments, recognise limitations, evaluate evidence, and make independent decisions.

AI assistance may affect these two functions differently. By providing feedback, suggesting alternatives, supporting technically demanding operations, AI can complement existing knowledge and help researchers to engage with tasks that would otherwise be too difficult or time-consuming. At the same time, when AI performs substantial parts of work, it may substitute for some of the cognitive effort through which competencies are developed. Immediate improvements in speed or output quality should therefore not automatically be interpreted as improvements in learning. Evidence from educational research similarly suggests that general-purpose generative AI can improve task performance without producing equivalent gains in knowledge or independent performance, particularly when its use is not guided by an explicit pedagogical purpose \citep{oecd2026}. Recent experimental studies provide more direct evidence of this distinction. \citet{fan2025} find that university students supported by ChatGPT improved their performance on a writing task without corresponding gains in knowledge acquisition, while also exhibiting less engagement in some metacognitive processes. Similarly, \citet{bastani} show that access to a general-purpose GPT interface improved performance during assisted practice but reduced subsequent unassisted performance, an effect which was largely mitigated when the AI tutor incorporated pedagogical guardrails.

The relevant distinction may then concern the tasks being delegated to AI as well as the way delegation is structured. The same activity may involve very different forms of human-AI interaction: AI can support the researcher's reasoning by providing feedback or supporting intermediate steps, but it can also replace some of the cognitive operations through which learning occurs. From this perspective, task-specific delegation and the mode of delegation are distinct: skill formation may depend on the research activity in which AI is used and on which parts of that activity remain under the researcher's cognitive control.

This distinction is particularly relevant in doctoral education. Recent contributions describe PhD researchers using AI for literature search, drafting, summarisation, coding, and other components of the research process, while also documenting concerns that extensive reliance on these tools could weaken some of the skills that doctoral training is intended to develop \citep{nordling2025}. Related arguments suggest that doctoral programmes will need to reconsider how researchers are trained as AI becomes increasingly capable of performing writing and data-analysis tasks \citep{sengupta2025}. The resulting challenge is not simply to determine whether AI should be permitted, but to distinguish between uses that support the development of expertise from uses that replace the exercise through which expertise is acquired.

Notice that our results cannot establish whether concerns about learning or skill preservation explain the observed attitudinal profiles. Nevertheless, the dominant \textit{``division of labour''} pattern is consistent with a distinction between activities that researchers are relatively willing to delegate and activities that remain more closely associated with the exercise of scientific judgement and intellectual responsibility. Literature tracking and summarisation may be regarded primarily as forms of support, whereas writing, data analysis, and experimental design may be perceived as more central to both scientific contribution and the development of research expertise. This interpretation remains tentative, since reported comfort may also reflect familiarity, access, perceived reliability, or other considerations unrelated to learning.

\subsection{Diffusion, normalisation, and the future of scientific work}
The profiles identified in the analysis should not be necessarily interpreted as permanent orientations. Norms surrounding emerging technologies may change as tools become more capable, accessible, or embedded in familiar practice. Uses that are initially regarded as inappropriate may become normalised; conversely, new knowledge about the impact of technology use may strengthen the previously weak boundaries. The task-specific patterns we report therefore provide a snapshot of a likely evolving normative environment. This evolution may be shaped by the interaction between individual attitudes and collective patterns of adoption. Researchers do not make decisions about AI use in isolation: they likely observe the practices of colleagues, respond to disciplinary and institutional expectations, and work in environments in which the adoption of particular tools may create new standards of speed, productivity, or output quality. Social influence is already recognized as an important dimension of technology adoption \citep{venkatesh2021}, and recent evidence suggests that it is also associated with researchers' use of AI \citep{acostaenriquez2025academia}. As particular applications become more common, researchers may revise their own evaluations of them or perceive increasing professional pressure to adopt them in order to remain aligned with prevailing practices.

This possibility does not imply that scientific communities will necessarily converge on uniformly positive attitudes towards AI. Normalisation may instead reinforce a differentiated division of labour in which some forms of assistance become routine, while others remain closely related to human action. The dominant profile we find may be interpreted as one possible configuration of such selective normalisation: AI is more readily accepted (now) for literature-related support than for writing, data analysis, or experimental design. At the same time, the existence of the other three profiles in our findings suggests that various directions of normative change are possible.

Institutional rules may both respond to and shape this process. Disclosure requirements, guidelines, publisher policies, evaluation practices can formalise distinctions between acceptable and unacceptable forms of assistance. These rules can in turn shape the normalisation process, through a reciprocal process in which individual attitudes, collective practices, and institutional governance evolve together.

The position of PhD researchers is particularly relevant in this respect. These researchers will become future supervisors, editors, reviewers, and research-governance decision-makers. The attitudes they develop during their training may influence not only their own future practices, but also what they transmit to subsequent cohorts and the way they evaluate the work of others. The profiles reported can then be interpreted as evidence of the normative orientations through which the next generation of researchers is choosing the place of AI in scientific work.

\subsection{Limitations and future research}
Several limitations qualify the interpretation of our findings. First, the survey items measure respondents' comfort with using AI rather than directly asking whether particular uses are legitimate, ethical, or appropriate. Comfort may reflect normative evaluations, but also familiarity with AI tools, access, confidence in using them, and beliefs about their reliability. The identified profiles should therefore be interpreted as configurations of task-specific attitudes with a normative dimension, rather than a direct measures of shared scientific norms. Future research could distinguish these mechanisms by combining measures of comfort with direct questions about legitimacy, authorship, disclosure, perceived responsibility, actual AI use, and technological competence.

Second, the data are cross-sectional and come from a self-selected international sample of PhD researchers. The observed associations cannot then be interpreted causally, nor can the analysis determine whether the identified profiles are stable. The sample also does not support claims about the prevalence of these profiles among the wider population of PhD researchers. Replication across disciplines and institutional settings, together with longitudinal research following the same researchers over time, would help establish how attitudes change their experience and peer practices.

Finally, our data do not connect task-specific attitudes to learning or scientific outcomes. Collecting information about the actual use of AI within each task may help determine whether greater acceptance of AI is associated with higher productivity, stronger (or weaker) skill formation, and originality of scientific work.

\subsection*{Data statement}

The data used in this article are publicly available at: \\
\url{https://figshare.com/articles/dataset/Nature_s_Graduate_Survey_2025/30084739}.

\subsection*{Acknowledgements}
The authors thank Massimiliano Castellani for his comments on a previous version of the manuscript. The paper also benefited from feedback from participants at the ``AI for Science and Innovation'' workshop, held at IMT School for Advanced Studies Lucca in April 2026.

\bibliographystyle{plainnat}
\bibliography{biblio}

\clearpage

\begin{table}[!ht]
    \small
    \centering
    \begin{tabular}{|l|rr|}
        \hline
        Item & Absolute frequency & Relative frequency \\
        \hline
        Writing research article &  &  \\
        - \emph{very comfortable} & 384 & 0.10 \\
        - \emph{somewhat comfortable} & 790 & 0.21 \\
        - \emph{neither} & 530 & 0.14 \\
        - \emph{somewhat uncomfortable} & 674 & 0.18 \\
        - \emph{very uncomfortable} & 1218 & 0.32 \\
        - \emph{don't know} & 189 & 0.05 \\
         &  &  \\
        Data collection \& analysis &  &  \\
        - \emph{very comfortable} & 383 & 0.10 \\
        - \emph{somewhat comfortable} & 733 & 0.19 \\
        - \emph{neither} & 533 & 0.14 \\
        - \emph{somewhat uncomfortable} & 710 & 0.19 \\
        - \emph{very uncomfortable} & 1180 & 0.31 \\
        - \emph{don't know} & 246 & 0.06 \\
         &  &  \\
        Designing experiments &  &  \\
        - \emph{very comfortable} & 347 & 0.09 \\
        - \emph{somewhat comfortable} & 806 & 0.21 \\
        - \emph{neither} & 611 & 0.16 \\
        - \emph{somewhat uncomfortable} & 685 & 0.18 \\
        - \emph{very uncomfortable} & 1027 & 0.27 \\
        - \emph{don't know} & 309 & 0.08 \\
         &  &  \\
        Tracking literature &  &  \\
        - \emph{very comfortable} & 589 & 0.16 \\
        - \emph{somewhat comfortable} & 1043 & 0.28 \\
        - \emph{neither} & 585 & 0.15 \\
        - \emph{somewhat uncomfortable} & 607 & 0.16 \\
        - \emph{very uncomfortable} & 716 & 0.19 \\
        - \emph{don't know} & 245 & 0.06 \\
         &  &  \\
        Summarising literature &  &  \\
        - \emph{very comfortable} & 683 & 0.18 \\
        - \emph{somewhat comfortable} & 1268 & 0.34 \\
        - \emph{neither} & 487 & 0.13 \\
        - \emph{somewhat uncomfortable} & 569 & 0.15 \\
        - \emph{very uncomfortable} & 612 & 0.16 \\
        - \emph{don't know} & 166 & 0.04 \\
        \hline 
    \end{tabular}
    \caption{Absolute and relative frequency distributions of the item responses. The relative frequency of a particular response category refers to its proportion within the corresponding item. Due to rounding, the relative frequencies do not sum to 1 for all the items. The frequencies are based on the full sample of 3785 respondents, which contains no missing values for these items.}
    \label{tab:summary_statistics_items}
\end{table}

\clearpage

\begin{table}[!ht]
    \small
    \centering
    \begin{tabular}{|l|rr|}
        \hline
        Covariate & Absolute frequency & Relative frequency \\
        \hline
        STEM researcher & 2828 & 0.76 \\
         &  &  \\
        First two years of PhD & 1418 & 0.38 \\
         &  &  \\
        Full-time researcher & 3247 & 0.87 \\
         &  &  \\
        AI use frequency &  &  \\
        - \emph{None} & 598 & 0.16 \\
        - \emph{Monthly or yearly} & 1137 & 0.31 \\
        - \emph{Daily or weekly} & 1987 & 0.53 \\
         &  &  \\
        34 years or younger & 2813 & 0.76 \\
         &  &  \\
        Male & 2024 & 0.54 \\
         &  &  \\
        Native English speaker & 542 & 0.15 \\
        \hline 
    \end{tabular}
    \caption{Absolute and relative frequency distributions for the covariates. The relative frequencies for the response categories of AI use frequency refer to their respective proportions within that item. The frequencies are based on the sample of 3722 respondents that have no missing values for any of the covariates.}
    \label{tab:summary_statistics_covariates}
\end{table}

\clearpage

\begin{table}[!ht]
    \centering
    \begin{tabular}{|l|r|}
        \hline
        Number of classes & ICLbic \\
        \hline
        1. & 97649 \\
        2. & 61138 \\
        3. & 59860 \\
        4. & 57453 \\
        5. & 57713 \\
        \hline
    \end{tabular}
    \caption{The integrated complete-data likelihood approximation of the Bayesian information criterion (ICLbic) for fitted standard latent class models with 1-5 classes. Lower values correspond to better goodness-of-fit. Each of the fitted models is based on the full sample of 3785 respondents.}
    \label{tab:enumeration}
\end{table}

\clearpage

\begin{table}[!ht]
    \small
    \centering
    \begin{tabular}{|l|rrrr|}
        \hline
         & Class 1 (0.44) & Class 2 (0.34) & Class 3 (0.16) & Class 4 (0.07) \\
         & \emph{``division of labour''} & \emph{``status quo''} & \emph{``all-purpose''} & \emph{``undecided''} \\
        \hline
        Writing research article &  &  &  &  \\
        - \emph{very comfortable} & 0.03 & 0.01 & 0.51 & 0.04 \\
        - \emph{somewhat comfortable} & 0.30 & 0.07 & 0.26 & 0.17 \\
        - \emph{neither} & 0.25 & 0.06 & 0.05 & 0.05 \\
        - \emph{somewhat uncomfortable} & 0.28 & 0.13 & 0.07 & 0.02 \\
        - \emph{very uncomfortable} & 0.14 & 0.71 & 0.09 & 0.08 \\
        - \emph{don't know} & 0.01 & 0.00 & 0.02 & 0.63 \\
         &  &  &  &  \\
        Data collection \& analysis &  &  &  &  \\
        - \emph{very comfortable} & 0.03 & 0.02 & 0.51 & 0.01 \\
        - \emph{somewhat comfortable} & 0.30 & 0.06 & 0.23 & 0.09 \\
        - \emph{neither} & 0.25 & 0.04 & 0.09 & 0.04 \\
        - \emph{somewhat uncomfortable} & 0.30 & 0.12 & 0.08 & 0.02 \\
        - \emph{very uncomfortable} & 0.10 & 0.76 & 0.06 & 0.04 \\
        - \emph{don't know} & 0.01 & 0.01 & 0.02 & 0.80 \\
         &  &  &  &  \\
        Designing experiments &  &  &  &  \\
        - \emph{very comfortable} & 0.03 & 0.02 & 0.47 & 0.02 \\
        - \emph{somewhat comfortable} & 0.33 & 0.05 & 0.29 & 0.08 \\
        - \emph{neither} & 0.28 & 0.06 & 0.10 & 0.04 \\
        - \emph{somewhat uncomfortable} & 0.27 & 0.15 & 0.06 & 0.04 \\
        - \emph{very uncomfortable} & 0.06 & 0.71 & 0.03 & 0.02 \\
        - \emph{don't know} & 0.03 & 0.01 & 0.05 & 0.81 \\
         &  &  &  &  \\
        Tracking literature &  &  &  &  \\
        - \emph{very comfortable} & 0.07 & 0.05 & 0.65 & 0.04 \\
        - \emph{somewhat comfortable} & 0.41 & 0.17 & 0.19 & 0.15 \\
        - \emph{neither} & 0.26 & 0.09 & 0.04 & 0.07 \\
        - \emph{somewhat uncomfortable} & 0.20 & 0.18 & 0.04 & 0.03 \\
        - \emph{very uncomfortable} & 0.05 & 0.47 & 0.05 & 0.03 \\
        - \emph{don't know} & 0.01 & 0.03 & 0.02 & 0.68 \\
         &  &  &  &  \\
        Summarising literature &  &  &  &  \\
        - \emph{very comfortable} & 0.10 & 0.04 & 0.76 & 0.05 \\
        - \emph{somewhat comfortable} & 0.50 & 0.21 & 0.18 & 0.24 \\
        - \emph{neither} & 0.22 & 0.08 & 0.01 & 0.09 \\
        - \emph{somewhat uncomfortable} & 0.16 & 0.22 & 0.01 & 0.01 \\
        - \emph{very uncomfortable} & 0.01 & 0.44 & 0.03 & 0.03 \\
        - \emph{don't know} & 0.00 & 0.01 & 0.01 & 0.57 \\
        \hline
    \end{tabular}
    \caption{The fitted 4-class model: class proportions in parenthesis in the table header, and conditional item-response probabilities in the body of the table. All proportions do not sum to 1 due to rounding. The estimation is based on the full sample of 3785 respondents.}
    \label{tab:fit_measurement_model}
\end{table}

\clearpage

\begin{table}[!ht]
    \small
    \centering
    \begin{tabular}{|l|rrr|}
        \hline
         & Class 2 & Class 3 & Class 4 \\
         & \emph{``status quo''} & \emph{``all-purpose''} & \emph{``undecided''} \\
        \hline
        STEM researcher & 0.307$^{**}$ & 0.132 & -0.237 \\
         & (0.11) & (0.13) & (0.18) \\
         &  &  &  \\
        First two years of PhD & -0.057 & 0.019 & -0.083 \\
         & (0.10) & (0.12) & (0.17) \\
         &  &  &  \\
        Full-time researcher & 0.262 & -0.249 & 0.032 \\
         & (0.15) & (0.17) & (0.23) \\
         &  &  &  \\
        AI use frequency &  &  &  \\
        (- \emph{None}) &  &  &  \\
        - \emph{Monthly or yearly} & -1.128$^{**}$ & -0.561$^*$ & -1.987$^{**}$ \\
         & (0.16) & (0.25) & (0.24) \\
        - \emph{Daily or weekly} & -1.911$^{**}$ & 0.434 & -2.402$^{**}$ \\
         & (0.16) & (0.23) & (0.24) \\
         &  &  &  \\
        34 years or younger & 0.297$^*$ & -0.150 & -0.340 \\
         & (0.12) & (0.14) & (0.19) \\
         &  &  &  \\
        Male & -0.413$^{**}$ & 0.272$^*$ & -0.009 \\
         & (0.10) & (0.12) & (0.17) \\
         &  &  &  \\
        Native English speaker & 1.047$^{**}$ & -0.777$^{**}$ & -0.462 \\
         & (0.14) & (0.26) & (0.35) \\
         &  &  &  \\
        Intercept & 0.418$^*$ & -1.052$^{**}$ & 0.290 \\
         & (0.21) & (0.29) & (0.29) \\
        \hline
    \end{tabular}
    \caption{The fitted model between the latent classes and the covariates: logistic parameters and their standard errors in parenthesis. The estimation is based on the 3722 observations with no missing values for the covariates. $^*p<0.05$. $^{**}p<0.01$.}
    \label{tab:fit_structural_model}
\end{table}
\end{document}